\documentclass[sigconf]{acmart}
\AtBeginDocument{%
  }

\copyrightyear{2026}
\acmYear{2026}
\setcopyright{cc}
\setcctype{by-nc-nd}
\acmConference[ICCAD '26]{IEEE/ACM International Conference on Computer-Aided Design}{November 08--12, 2026}{San Jose, CA, USA}
\acmBooktitle{IEEE/ACM International Conference on Computer-Aided Design (ICCAD '26), November 08--12, 2026, San Jose, CA, USA}
\acmDOI{10.1145/3831252.3834213}
\acmISBN{979-8-4007-2873-0/2026/11}

\usepackage{booktabs}
\usepackage{multirow}
\usepackage{pifont}

\begin{document}

\title{PolyCIM: Improving Data Reuse in Digital CIM Accelerators with Polyhedral-Based Compilation}

\titlenote{
This work was supported by the National Natural Science Foundation of China (Grant No. 62572036).
The corresponding author is \textit{Jianlei Yang} (\url{jianlei@buaa.edu.cn}).
}

\author{Yingjie Qi}
\authornote{Both authors contributed equally to this research.}
\affiliation{
  \institution{Beihang University}
  \city{Beijing}
  \country{China}
}

\author{Cenlin Duan}
\authornotemark[2]
\affiliation{
  \institution{Beihang University}
  \city{Beijing}
  \country{China}
}

\author{Yiou Wang}
\affiliation{
  \institution{Beihang University}
  \city{Beijing}
  \country{China}
}

\author{Yikun Wang}
\affiliation{
  \institution{Beihang University}
  \city{Beijing}
  \country{China}
}

\author{Xiaolin He}
\affiliation{
  \institution{Beihang University}
  \city{Beijing}
  \country{China}
}

\author{Weisheng Zhao}
\affiliation{
  \institution{Beihang University}
  \city{Beijing}
  \country{China}
}

\author{Jianlei Yang}
\affiliation{
  \institution{Beihang University}
  \city{Beijing}
  \country{China}
}

\begin{abstract}
Digital Compute-in-Memory (CIM) presents a promising solution for accelerating deep neural networks (DNNs) through the integration of computational logic directly within memory arrays.
However, mapping modern DNN operators to CIM accelerators often results in severe array underutilization, due to the strict data reuse constraints imposed by the rigid CIM array structure.
We observe that data reuse in modern DNNs forms hyperplane structures often oriented along non-axial directions, rendering them invisible to conventional mapping methods that only exploit axis-aligned reuse.
In this work, we propose PolyCIM, a polyhedral-based compilation framework for CIM architectures that systematically exposes and realigns these hyperplanes through affine transformations.
PolyCIM provides a unified abstraction capable of efficiently representing both diverse DNN workloads and digital CIM architectures.
Through data reuse exposure, computation mapping, and data movement optimization, PolyCIM generates mappings for CIM architectures that achieve superior array utilization and performance.
Experimental results show that PolyCIM delivers up to $4\times$ improvement in macro utilization and $3.2\times$ speedup, effectively bridging the gap between modern DNN operators and CIM architectures.
\end{abstract}



\maketitle

\section{Introduction}

The continuous scaling of deep neural networks (DNNs) has made data movement between memory and compute the primary performance bottleneck in conventional architectures.
To address this `Memory Wall' challenge, Compute-in-Memory (CIM) architectures have emerged as a promising paradigm by performing matrix-vector multiplication (MVM) operations directly inside the memory array.
Most early CIM designs are analog CIM~\cite{shafiee-2016-isaac, ankit-2019-puma, biswas-2018-convram}, which perform multiply-accumulate (MAC) operations in the voltage or current domain to achieve high efficiency through device-level parallelism.
However, their reliance on analog-to-digital converters (ADCs) and sensitivity to analog non-idealities constrain both the number of simultaneously activated rows and columns and the overall precision of the computation.
In contrast, digital CIM~\cite{chih-2021-sram, yan-2022-sram, duan-2024-ddc, duan-2024-sparsity} performs the entire computation in the digital domain, integrating MAC logic directly into the memory array.
This approach eliminates analog non-idealities and ADC overhead, offering more robust computation and higher parallelism that make it well-suited for accelerating modern DNN workloads.

\begin{figure}[t]
\centering
\includegraphics[width=\linewidth]{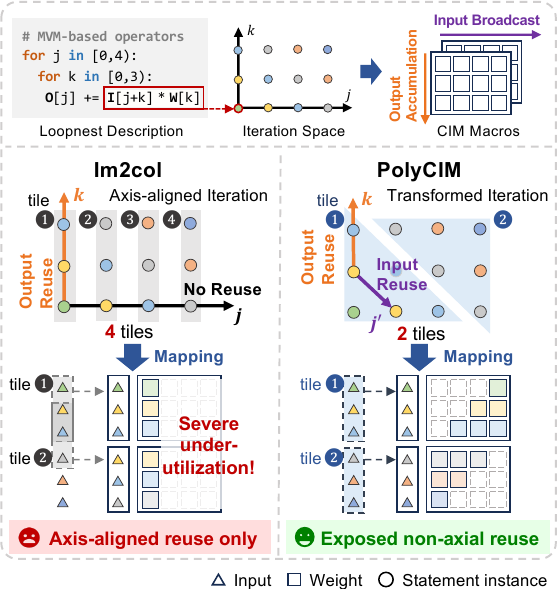}
\caption{Differences between im2col and PolyCIM mapping on CIM, illustrated using an example of a Conv1D operator.}
\Description{Comparison of two Conv1D mappings. Im2col divides the axis-aligned iteration space into four tiles that leave many CIM macro cells idle. PolyCIM transforms the iteration space into two tiles that expose non-axial input and output reuse and occupy more macro cells.}
\label{fig:Compare-im2col-PolyCIM}
\end{figure}

However, as the field of DNNs evolves towards more efficient architectures~\cite{tan-2019-efficientnet, xie-2017-resnext, mehta-2019-espnetv2, shamsafar-2022-mobilestereonet}, mapping modern DNN operators to rigid CIM array structures becomes increasingly challenging.
The structural constraints of a digital CIM array require each row to be activated atomically.
Every memory cell along the row performs its multiplication in parallel, and the resulting partial products accumulate down each column.
The image-to-column (im2col) method, adopted by most CIM designs, aligns with this structure by unrolling weight filters into array columns and broadcasting input activations along rows. 
While effective for mapping standard convolutions, this approach ties array utilization directly to the number of filters that share the same input activations.
When this number falls below the array column count, the unused columns still activate under the rigid CIM structure but produce no useful output, leading to severe underutilization.
This presents significant limitations for efficiently supporting emerging operators on CIM, such as depthwise and grouped convolutions.

Fig.~\ref{fig:Compare-im2col-PolyCIM} illustrates this mapping challenge with a minimal example, a one-dimensional convolution (Conv1D) operator containing a single filter.
With only one filter to map per input, im2col can only meaningfully utilize one column while still activating the entire array.
The fundamental problem is that the CIM array only supports data reuse that runs along its rows or columns.
Since mappings such as im2col apply a fixed and operator-independent transformation, they can expose only those reuse patterns that already align with these two directions.
However, for modern DNN operators, iterations accessing the same data element form hyperplane structures that often lie along non-axial directions in the iteration space, leaving these reuse opportunities beyond the reach of such mappings.
Since DNN operand accesses are affine functions of loop indices, these hyperplanes can be systematically realigned with the array axes through affine transformations automatically derived for each operator, converting hidden reuse patterns into exploitable parallelism.

Despite recent efforts in improving CIM utilization, existing approaches lack the mathematical framework to systematically expose and exploit these non-axial reuse patterns.
Specialized solutions~\cite{zhang-2021-sdk, rhe-2022-vwsdk} rely on manual pattern identification and transformation, while compilation and dataflow optimization frameworks for CIM accelerators~\cite{siemieniuk-2022-occ, sun-2023-pimcomp, qi-2025-cimflow, drebes-2020-tccim, han-2022-polyhedral, qi-2026-ciminus, he-2026-miredo} remain confined to axis-aligned transformations.
To bridge this gap, we propose PolyCIM, a polyhedral-based compilation framework for digital CIM architectures.
Given a DNN operator, PolyCIM automatically derives an affine transformation that exposes the non-axial reuse and aligns it with the array axes, which yields substantial gains in macro utilization and performance for emerging operators that im2col-like mappings cannot handle efficiently.
The contributions of this paper can be summarized as follows:
\begin{itemize}
\setlength{\emergencystretch}{1.5em}
\item We formulate the exposure of non-axial data reuse in CIM as a polyhedral transformation problem, expanding the class of reuse patterns that CIM hardware can exploit beyond what conventional mappings can capture.
\item We propose PolyCIM, a polyhedral-based compilation framework that automatically transforms DNN operators into high-utilization CIM mappings through data reuse exposure, computation mapping, and data movement optimization.
\item Extensive evaluation demonstrates that compared to conventional mapping methods, PolyCIM achieves up to $4\times$ macro utilization improvement and $3.2\times$ speedup across diverse DNN workloads and CIM configurations.
\end{itemize}

\section{Background}\label{sec:background}

\subsection{DNN Mappings on CIM Architectures}
Mapping DNN workloads to CIM architectures requires transforming DNN operators into MVM operations that align with the CIM structural constraints.
To support the increasing diversity of DNN operators, dedicated compilation frameworks such as OCC~\cite{siemieniuk-2022-occ}, PIMCOMP~\cite{sun-2023-pimcomp}, and CIMFlow~\cite{qi-2025-cimflow} aim to achieve better flexibility through multi-level intermediate representation (IR) transformations.
Meanwhile, frameworks such as TC-CIM~\cite{drebes-2020-tccim} and Han et al.~\cite{han-2022-polyhedral} employ polyhedral techniques for loop optimization and operator detection.
Their compilation flows lower the identified computations to fixed matrix multiplication primitives, whose array mappings are also aligned with the hardware axes.

Most of these frameworks rely on the im2col transformation, which reshapes the operands of a convolution so that the CIM array can execute it as a dense MVM operation.
Each output filter is unrolled into a dedicated column of the array, and the input activations are reorganized into a sequence of sliding windows, where each window holds the input values that a filter consumes at one output position.
A single cycle of the array then broadcasts one window across all rows and produces one output per active column, yielding every output channel for that position in parallel.
However, this fixed layout ties array utilization directly to the number of filters that share the same input window, leaving many columns idle whenever that number is small.

A few specialized methods have attempted to improve CIM utilization by targeting specific non-axial reuse patterns.
The Shifted and Duplicated Kernel (SDK) family of approaches~\cite{zhang-2021-sdk, rhe-2022-vwsdk} exploits the sliding-window reuse of 2D convolutions by loading several overlapping input windows into the rows of the array and placing shifted copies of the same filter in neighboring columns, so that each column computes a different output position while drawing from a shared pool of input activations.
This technique noticeably improves utilization for standard 2D convolutions.
However, its transformation space is manually designed around the sliding-window pattern of a single operator class, and the shift-and-duplicate layout does not generalize to operators whose reuse geometry differs.
Moreover, SDK was developed for analog RRAM-based CIM and relies on spare array capacity to place shifted copies of a kernel.
The limited array dimensions of digital SRAM CIM constrain such duplication, making the resulting mapping equivalent to im2col when the array cannot accommodate additional shifted kernel copies.

Despite these efforts, no existing framework offers a principled and generalizable way to expose the non-axial reuse patterns found in modern DNN operators.
This gap motivates the polyhedral-based approach of PolyCIM, whose underlying model we introduce in the next subsection.

\subsection{Polyhedral Model}

The polyhedral model is a mathematical framework for analyzing and transforming loop-nested programs with affine loop bounds and array accesses~\cite{bondhugula-2008-pluto}.
By representing a loop nest as an integer polyhedron, it turns dependence analysis and loop transformations into exact algebraic operations, with the legality of each transformation provable directly from the polyhedral form.
In this framework, an iteration vector $\mathbf{i}_S = [i_1, \ldots, i_n]$ represents one execution instance of a statement $S$, composed of all enclosing loop indices.
The iteration domain $\mathcal{D} = \{\mathbf{i}_S \mid q(\mathbf{i}_S) \leq \mathbf{0}\}$ collects all valid iteration vectors as the integer points inside a polyhedron, where $q$ is an affine function encoding loop bounds.
Memory accesses from within the loop nest are described by access relations $\mathcal{A}_M = \{\mathbf{i}_S \rightarrow M[\mathbf{j}] \mid q(\mathbf{i}_S, \mathbf{j}) = \mathbf{0}\}$, which map each iteration to the element it touches in operand array $M$.
When the array indices are affine functions of the loop variables, each relation $\mathcal{A}_M$ is fully specified by an access matrix $\mathbf{A}_M$ and a constant offset $\mathbf{c}_M$ via the linear formula
\begin{equation}
\mathbf{j} = \mathbf{A}_M \mathbf{i}_S + \mathbf{c}_M,
\end{equation}
where $\mathbf{A}_M$ has one row per array dimension expressing its linear dependence on the loop indices.
Two iterations $\mathbf{i}_1$ and $\mathbf{i}_2$ then access the same element of $M$ exactly when $\mathbf{A}_M(\mathbf{i}_1 - \mathbf{i}_2) = \mathbf{0}$, so the set of iterations sharing any fixed element forms an affine subspace of the iteration domain whose direction vectors lie in the null space of $\mathbf{A}_M$.
These affine subspaces are the reuse hyperplanes that polyhedral compilation can expose.

Program transformations in the polyhedral model are expressed as schedules $\mathcal{S} = \{\mathbf{i}_S \rightarrow \mathbf{t} \mid q(\mathbf{i}_S, \mathbf{t}) = \mathbf{0}\}$, which assign an execution timestamp $\mathbf{t}$ to each iteration.
The iteration domain and access relations fix what each statement reads and writes, while the schedule is free to reshape the execution order into any form that respects the data dependences.
This separation allows the compiler to search over a rich space of schedules while reusing a single dependence analysis.
A schedule can equivalently be written as an affine map
\begin{equation}
\mathbf{t} = \mathbf{T}\mathbf{i}_S + \mathbf{c},
\end{equation}
where the schedule matrix $\mathbf{T}$ has one row per time dimension specifying how iterations are ordered along it, and $\mathbf{c}$ is a constant offset.
Classical loop transformations then correspond to particular choices of $\mathbf{T}$, including a permutation matrix for interchange, a unimodular matrix for skewing or reversal, and a division into outer and inner bands for tiling.
Composition under matrix multiplication lets these primitives be combined into richer reshapings of the iteration space, provided the resulting schedule respects the data dependences captured in the polyhedral representation.
For CIM architectures, these transformation capabilities enable the alignment of complex data reuse patterns with the rigid structure of memory arrays.
A single affine schedule can rotate the iteration basis so that the reuse hyperplanes of an operand become coordinate directions, turning reuse that conventional axis-aligned approaches cannot see into parallelism that the hardware can directly exploit.

\section{PolyCIM Framework}\label{sec:methodology}

\subsection{Framework Overview}\label{sec:Overview}

\begin{figure}[t]
    \centering
    \includegraphics[width=\linewidth]{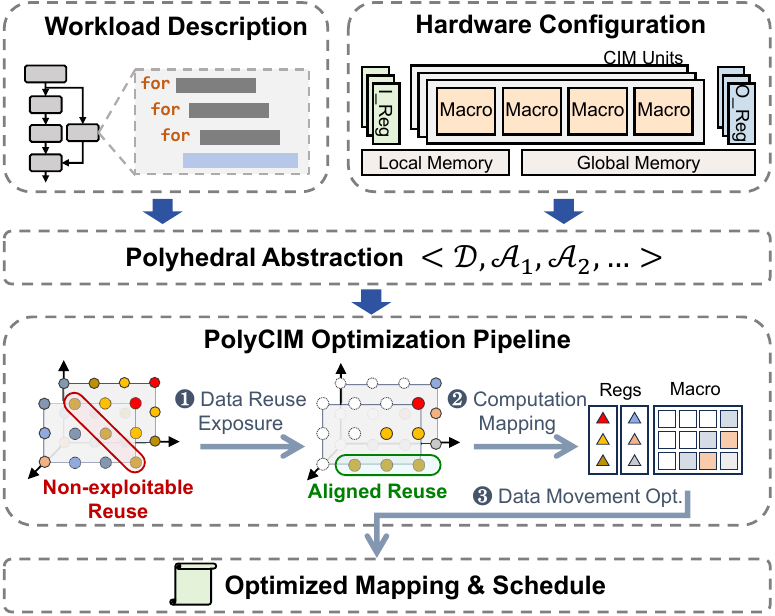}
    \caption{Overview of the PolyCIM framework.}
    \Description{A workload description and a CIM hardware configuration are converted into a common polyhedral abstraction. The PolyCIM pipeline then exposes data reuse, maps computations to registers and macros, and optimizes data movement to produce an optimized mapping and schedule.}
    \label{fig:overview}
\end{figure}

\begin{figure*}
    \centering
    \includegraphics[width=\textwidth]{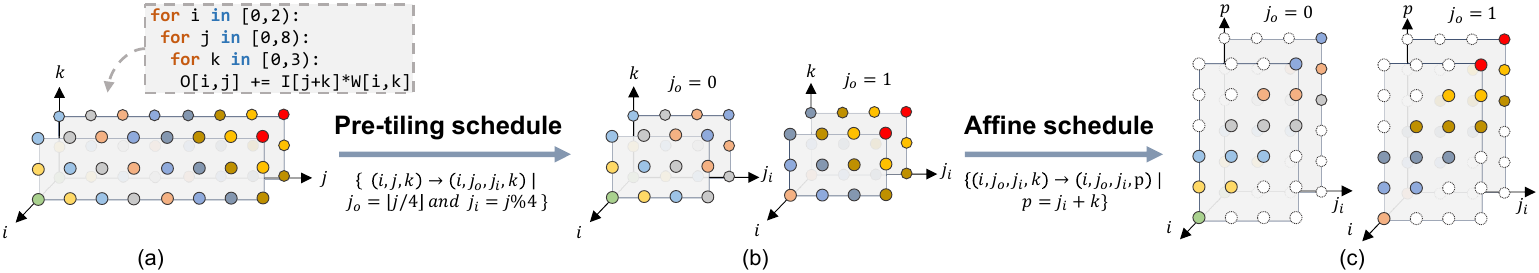}
    \caption{Example of data reuse exposure with a Conv1D operator. Same-colored iteration points access the same input data.}
    \Description{A Conv1D loop nest is shown as a three-dimensional iteration space. Pre-tiling decomposes one iteration axis, and an affine schedule then aligns same-colored points that access the same input along regular directions in the transformed space.}
    \label{fig:reuse-example}
\end{figure*}

PolyCIM is a compilation framework that transforms DNN operators into optimized mappings for CIM accelerators.
It leverages the insight that data reuse in DNNs naturally forms hyperplane structures in the iteration space, often along oblique directions that conventional axis-aligned methods cannot exploit.
As illustrated in Fig.~\ref{fig:overview}, the framework takes an operator description and a hardware configuration as input and constructs a common polyhedral representation, which is then refined through three successive stages before being lowered to executable code: data reuse exposure, computation mapping, and data movement optimization.

\textbf{Data reuse exposure} starts from the raw iteration space and its access relations and identifies the directions along which operands are reused.
Because these directions often lie oblique to the coordinate axes, PolyCIM first applies pre-tiling to decompose iteration axes into finer sub-coordinates, then performs affine scheduling to rotate the iteration basis so that reuse aligns with the new axes.
The result is a scheduled iteration space in which each axis carries the reuse of a specific operand.
Given this reuse-aligned space, \textbf{computation mapping} places iterations onto the macro grid in two phases.
A \textit{virtual mapping} first coalesces scattered reuse into unified row, column, and macro dimensions as if the macro were unbounded, after which a \textit{physical mapping} fits these dimensions within the actual macro bounds through a post-tiling schedule.
The inner levels of this loop nest map directly onto macro positions, while the outer levels drive inter-macro parallelism.
Finally, \textbf{data movement optimization} resolves how operands flow through the memory hierarchy.
PolyCIM selects a buffering level for each operand and reorders the loop nest so that accesses remain contiguous and redundant transfers between on-chip memory and macro-local registers are minimized.
The resulting schedule is handed to code generation, which emits the executable CIM program.

\subsection{PolyCIM Abstractions}

PolyCIM employs unified polyhedral abstractions to model both DNN operators and CIM architectures within the same mathematical framework.
PolyCIM primarily targets MVM-based operators, such as convolution and fully connected layers, that dominate DNN computation.
It represents each operator as a quadruple $\langle\mathcal{D},\allowbreak \mathcal{A}_{I},\allowbreak \mathcal{A}_{W},\allowbreak \mathcal{A}_{O}\rangle$, where $\mathcal{D}$ represents the iteration space and $\mathcal{A}_{I/W/O}$ capture memory access relations for inputs, weights, and outputs respectively.
This abstraction can represent diverse DNN operators including standard, depthwise (DWConv), grouped (GConv), and dilated (DILConv) convolutions.
To illustrate these abstractions, we use the one-dimensional convolution (Conv1D) depicted in Fig.~\ref{fig:reuse-example} as a running example throughout the methodology: an operator with two output channels, eight output positions per channel, and a kernel of width three.
Its iteration space and access relations are given by
\begin{equation}
\begin{aligned}
\mathcal{D} &= \{(i,j,k) \mid 0 \leq i < 2 \wedge 0 \leq j < 8 \wedge 0 \leq k < 3\}, \\
\mathcal{A}_I &= \{(i,j,k) \rightarrow I[j+k]\}, \\
\mathcal{A}_W &= \{(i,j,k) \rightarrow W[i,k]\}, \\
\mathcal{A}_O &= \{(i,j,k) \rightarrow O[i,j]\}.
\end{aligned}
\end{equation}

For hardware abstractions, PolyCIM models a representative CIM architecture consisting of CIM units, auxiliary compute units, and supporting memory hierarchies.
As shown in Fig.~\ref{fig:overview}, each CIM unit is composed of multiple macros with flexible inter-macro parallelization, unlike the rigid data reuse constraints within each macro.
Similar to the operator abstraction, a CIM unit is modeled as $\langle\mathcal{D}_{\text{CIM}},\allowbreak \mathcal{A}_{\text{in\_reg}},\allowbreak \mathcal{A}_{\text{mem}},\allowbreak \mathcal{A}_{\text{out\_reg}}\rangle$, where $\mathcal{D}_{\text{CIM}}$ indexes the computation space $(i_m, i_r, i_c)$ across macros, rows, and columns.
These three access relations encode the structural constraints that give each macro its rigidity.
Input broadcasting appears in $\mathcal{A}_{\text{in\_reg}}$, which indexes only $(i_m, i_r)$ so that every column in a row reads the same register.
Output accumulation appears symmetrically in $\mathcal{A}_{\text{out\_reg}}$, which indexes only $(i_m, i_c)$ so that every row in a column writes to the same register.
The local memory relation $\mathcal{A}_{\text{mem}}$, however, retains the full $(i_m, i_r, i_c)$ triple and maps each position to its own memory location.
This unified abstraction across software and hardware enables systematic transformation between operator iteration spaces and CIM computation spaces while preserving data dependencies and hardware constraints.

\subsection{Data Reuse Exposure}\label{subsection:explore-data-reuse}

For modern DNN operators, the fundamental challenge in CIM mapping is that their data reuse patterns often lie along oblique directions in the iteration space, which are invisible to conventional axis-aligned approaches. 
PolyCIM exposes these hidden opportunities through a two-phase transformation strategy: pre-tiling to control iteration space expansion, followed by affine scheduling to realign reuse patterns with CIM array structure. 
Fig.~\ref{fig:reuse-example} demonstrates this process on our Conv1D running example.
We formulate the optimization objective for a given operator as maximizing average macro utilization:
\begin{equation}
\text{util} = \frac{\text{MACs}}{P_\text{MAC} \times \text{cycles}},
\end{equation}
where MACs denotes total multiply-accumulate operations of the operator, $P_\text{MAC} = R\times C$ represents concurrent MAC operations per macro (with $R$ rows and $C$ columns that can be activated simultaneously), and cycles indicates required execution cycles.
Maximizing utilization equivalently minimizes execution cycles for a given operator and hardware configuration.
We now detail each transformation phase below.

\subsubsection{Pre-Tiling Schedule}\label{subsubsection:pre-tiling-schedule}

\begin{figure}
    \centering
  \includegraphics[width=\linewidth]{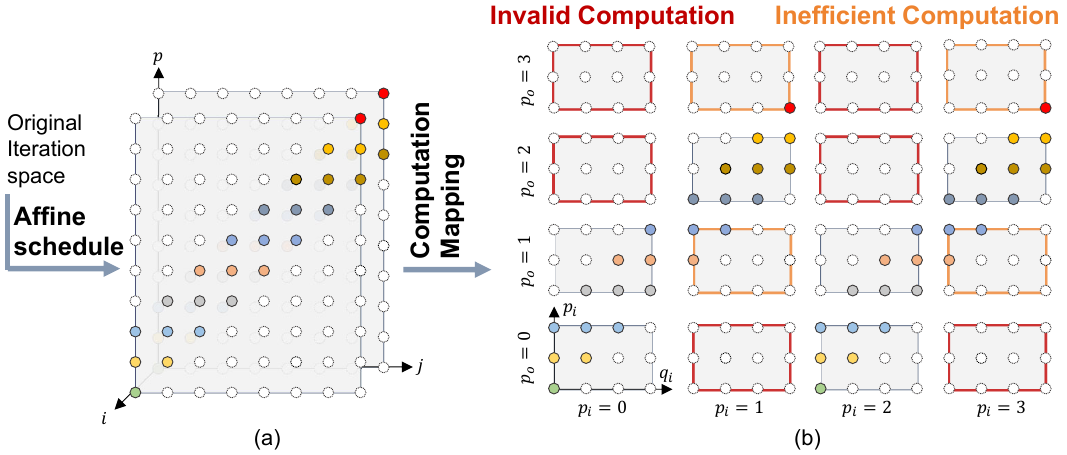}
  \caption{Example of the iteration space of a 1D convolution scheduled without pre-tiling.}
  \Description{An affine schedule expands the original Conv1D iteration space before computation mapping. The mapped grid contains red regions of invalid computation and orange regions of inefficient computation, illustrating the expansion that pre-tiling is designed to avoid.}
  \label{fig:pretile-example}
\end{figure}

While affine transformations effectively expose data reuse, they may also significantly expand the iteration space bounding box, introducing invalid iteration points that result in wasted CIM computation cycles.
Pre-tiling addresses this by strategically decomposing iteration dimensions before transformation, containing the expansion within manageable bounds.
As shown in Fig.~\ref{fig:pretile-example}, directly applying affine scheduling to the Conv1D example without pre-tiling expands the iteration space from its original $2 \times 8 \times 3$ to $2 \times 8 \times 10$, resulting in numerous underutilized or invalid CIM operations after computation mapping.
We now formalize the pre-tiling schedule and describe the heuristics that prune its candidate space.

\textbf{Schedule Construction}.
For each iteration axis $k$ with length $s_k$, pre-tiling decomposes $i_k$ into a sequence of sub-coordinates whose sizes form a factorization of $s_k$.
For $s_k>1$, the set of valid factorizations for axis $k$ is
\begin{equation}
\mathcal{F}_k = \{(p_1, \ldots, p_\ell) \mid p_1, \ldots, p_\ell \in \mathbb{Z}_{>1},\ p_1 \cdots p_\ell = s_k\},
\end{equation}
and for a unit-length axis, we define $\mathcal{F}_k=\{(1)\}$.
The full space of pre-tiling candidates across all $n$ axes is the Cartesian product 
\begin{equation}
    \mathcal{F} = \mathcal{F}_1 \times \cdots \times \mathcal{F}_n.
\end{equation}
Given a factorization $(p_1, \ldots, p_\ell) \in \mathcal{F}_k$, the pre-tiling schedule replaces $i_k$ with $\ell$ sub-coordinates $(i_k^{(1)}, \ldots, i_k^{(\ell)})$ related by the mixed-radix decomposition
\begin{equation}
i_k = \sum_{j=1}^\ell i_k^{(j)} \prod_{j' = j+1}^\ell p_{j'}, \quad 0 \leq i_k^{(j)} < p_j.
\end{equation}
Applying this decomposition across all axes yields the pre-tiling schedule $\mathcal{S}_\text{Pre}$, which PolyCIM applies to the operator before affine scheduling.

\textbf{Pruning Heuristics}.
The candidate space $\mathcal{F}$ grows combinatorially with the number of axes and their factorization counts, so the pre-tiling search applies two complementary pruning heuristics: axis exclusion and symmetry-based deduplication.
Axes that do not participate in affine transformations are excluded from the search, since their tile sizes cannot affect the subsequent scheduling outcome.
The symmetry heuristic targets convolutions with symmetric spatial dimensions, such as 2D or 3D convolutions with equal heights and widths.
For such operators, tile-size assignments that differ only by permuting symmetric axes yield identical downstream utilization, so only one representative from each equivalence class is retained.
With these restrictions in place, PolyCIM evaluates the remaining candidates and selects one that minimizes iteration space expansion during affine scheduling while preserving reuse opportunities.

\subsubsection{Affine Schedule} \label{subsubsection:affine-schedule}

Following pre-tiling, PolyCIM applies affine transformations to expose non-axial data reuse patterns.
While affine transformations can theoretically expose any linear reuse pattern, finding those that maximize CIM utilization requires exploring an enormous search space.
PolyCIM addresses this challenge by reformulating transformation discovery as a basis search problem.
Specifically, we identify basis vectors parallel to data reuse hyperplanes, then transform the iteration space based on these bases such that reuse occurs along the new coordinate axes.

\textbf{Reuse Hyperplanes}.
Iterations that access the same data element naturally form hyperplane structures within the iteration space. 
Consider an array $M$ with access relation $\mathcal{A}_M=\{ \mathbf{i} \rightarrow M[\mathbf{A}\mathbf{i}+\mathbf{b}]\}$, where $\mathbf{A} \in \mathbb{Z}^{m \times n}$ maps $n$-dimensional iteration vectors to $m$-dimensional array indices. 
The set of iterations accessing the same element forms a hyperplane:
\begin{equation}
\mathcal{H} = \{\mathbf{i} \in \mathcal{D} \mid \mathbf{A}\mathbf{i}=\mathbf{c}\},
\end{equation}
where $\mathbf{c}$ is a constant vector and $\mathbf{A}$ defines the hyperplane orientation.
These hyperplanes represent reuse opportunities, since iterations on the same hyperplane share data.
Our goal is to find transformations that align these hyperplanes with CIM macro dimensions.

\textbf{Basis Vector Constraints}.
Since standard basis vectors often cut across reuse hyperplanes, which prevents efficient data reuse, we construct new bases that align with these hyperplanes while satisfying the following two constraints:

\underline{\textit{\ding{182} Direction Constraint}}.
The reuse direction constraint refers to the requirement that basis vectors must lie parallel to reuse hyperplanes to ensure iterations along these directions access the same data.
Specifically, for an $n$-dimensional iteration space $\mathcal{D}$ and an $m$-dimensional array $M$, iteration vectors in the direction of basis $\mathbf{b} = [b_1, \ldots, b_n]^T $ can reuse the same element of $M$ only if $\mathbf{A}\mathbf{b}=\mathbf{0}$.

\underline{\textit{\ding{183} Degree Constraint}}.
When the reuse hyperplane spans multiple dimensions, multiple bases may satisfy the direction constraint, and we need an additional criterion to select the one exposing the most reuse.
Given such a basis $\mathbf{b}$, the direction constraint guarantees that stepping by $\mathbf{b}$ from any iteration lands on another iteration accessing the same element.
The set of iterations reachable this way from a starting point $\mathbf{i}$ is the reuse set
\begin{equation}
\mathcal{R}(\mathbf{i}, \mathbf{b}) = \{\mathbf{i} + k\mathbf{b} \mid k \in \mathbb{Z},\ \mathbf{i} + k\mathbf{b} \in \mathcal{D}\}.
\end{equation}
A basis is only useful if it exposes substantive reuse, so we require at least one starting point $\mathbf{i} \in \mathcal{D}$ from which $|\mathcal{R}(\mathbf{i}, \mathbf{b})| \geq r+1$, meaning that at least $r$ consecutive reuse steps along $\mathbf{b}$ stay within the domain.
For a rectangular $\mathcal{D}$ with axis sizes $\mathbf{m} = [m_1, \ldots, m_n]$, this condition is equivalent to
\begin{equation}
-\mathbf{m} < r \cdot \mathbf{b} < \mathbf{m},
\end{equation}
since each component $b_i$ must leave enough room along axis $i$ for $r$ consecutive reuse steps.
We adopt this as the reuse degree constraint.

\textbf{Basis Construction}.
We aim to select a set of basis vectors that maximizes reuse while satisfying both constraints.
We set $r = 1$ as the minimum degree threshold, guaranteeing at least one reuse step between two distinct iterations along each basis vector.
For each operand array $M \in \{I, W, O\}$, we construct a candidate basis set
\begin{equation}
\mathcal{B}_M = \{ \mathbf{b} \mid \mathbf{A}_M \mathbf{b} = \mathbf{0} \wedge -\mathbf{m} < \mathbf{b} < \mathbf{m} \}.
\end{equation}
We employ the Integer Set Library (ISL) to efficiently compute these sets.
From $\mathcal{B}_I \cup \mathcal{B}_W \cup \mathcal{B}_O$, we must select $n$ linearly independent vectors that span the iteration space, drawing at least one from $\mathcal{B}_I$ and at least one from $\mathcal{B}_O$ so that both input and output reuse are exposed.
Since multiple valid combinations may exist, we evaluate each through the subsequent optimization stages and retain the combination with the highest average macro utilization.

\textbf{Coordinate Transformation}.
Having selected $n$ linearly independent basis vectors, we construct the transformation matrix $\mathbf{B} = [\mathbf{b}_1, \ldots, \mathbf{b}_n]$ and compute its inverse $\mathbf{P} = \mathbf{B}^{-1}$ to obtain the 
coordinate transformation.
The nonsingularity of $\mathbf{B}$ implies that $\mathbf{P}$ is injective over $\mathcal{D}$, so the resulting schedule changes only the order in which products are accumulated for each output element.
Since $\mathbf{P}$ may contain fractional entries, we scale each row by the least common multiple (LCM) of its denominators to ensure integer coefficients, preserving transformation semantics while enabling efficient implementation.
Finally, the resulting affine scheduling can be represented as
\begin{equation}
\mathcal{S}_\text{Aff}=
\lbrace \mathbf{j} \rightarrow \mathbf{P} \mathbf{j} 
\mid \mathbf{j} \in \mathbb{Z}^n, \mathbf{P} \in \mathbb{Z}^{n \times n}
\rbrace,
\end{equation}
which realigns the iteration space such that data reuse occurs along the new coordinate axes, making reuse patterns directly exploitable by CIM macros.

\subsection{Computation Mapping}

\begin{figure}
  \centering
  \includegraphics[width=\linewidth]{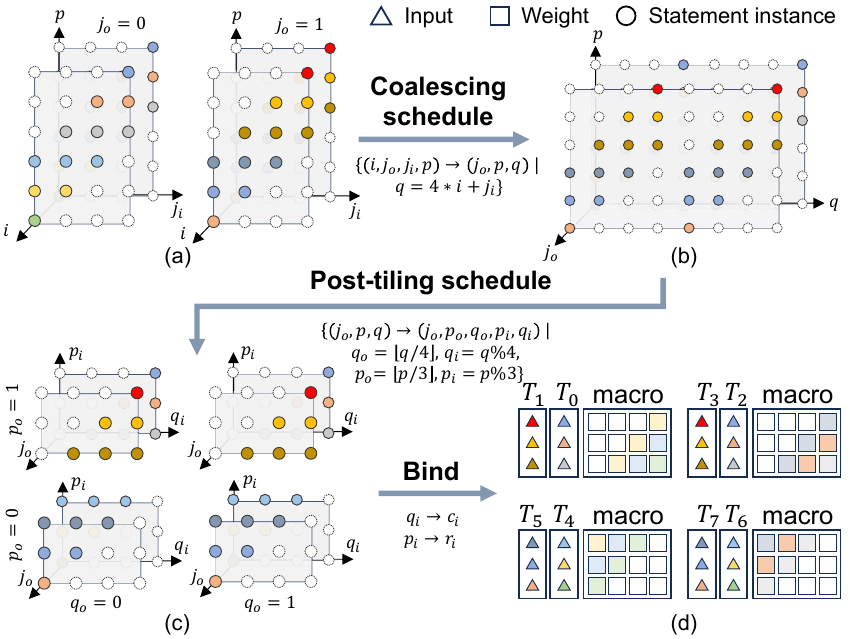}
  \caption{Example of computation mapping.}
  \Description{A transformed Conv1D iteration space is coalesced into unified dimensions, post-tiled into inner and outer coordinates, and bound to macro rows and columns. The outer tiles are then distributed across four CIM macros.}
  \label{fig:computation-mapping}
\end{figure}

After exposing data reuse along iteration space bases, PolyCIM must map the transformed computation onto the CIM architecture while respecting both the rigid constraints within each macro and the flexible parallelization opportunities across multiple macros.
Following affine scheduling, each iteration axis is labeled by the operand whose reuse it carries.
An \textit{input-reuse} axis has its basis in $\mathcal{B}_I$, so that iterations along it access the same input element.
\textit{Output-reuse} and \textit{weight-reuse} axes are defined analogously via $\mathcal{B}_O$ and $\mathcal{B}_W$, while an axis whose basis lies in none of these sets is a \textit{non-reuse} axis that iterates over distinct elements of every operand.
PolyCIM employs a hierarchical mapping strategy that first coalesces scattered reuse dimensions into unified axes matching macro structure, tiles them to fit physical macro dimensions, and finally distributes the remaining iterations across multiple macros to further exploit available parallelism.
Fig.~\ref{fig:computation-mapping} illustrates the macro-level mapping process, showing how a transformed Conv1D operator maps to CIM macros through coalescing, tiling, and binding.

\subsubsection{CIM Macro Mapping}

While affine transformation exposes data reuse along basis directions, the resulting iteration space rarely aligns with CIM macro constraints.
Specifically, data reuse may span multiple dimensions, whereas macros require exactly two: one for row-wise input broadcasting and one for column-wise output accumulation.
PolyCIM bridges this gap in two stages: a \textit{virtual mapping} stage that uses a coalescing schedule to merge scattered reuse into unified dimensions as if the macro were unbounded, followed by a \textit{physical mapping} stage that uses a post-tiling schedule to fit these dimensions within the physical macro bounds.

\textbf{Coalescing Schedule}.
Multiple axes reusing the same data must be merged into single dimensions matching macro structure.
For input-reuse axes $\{i_1^{\text{in}},...,i_p^{\text{in}}\}$ with sizes $\{s_1^{\text{in}},...,s_p^{\text{in}}\}$ and output-reuse axes $\{i_1^{\text{out}},...,i_q^{\text{out}}\}$ with sizes $\{s_1^{\text{out}},...,s_q^{\text{out}}\}$, the coalescing schedule creates unified dimensions
\begin{equation}
\mathcal{S}_{\text{Coal}}= \{ (i_1,...,i_n) \rightarrow (t_1,...,t_k, r, c) \},
\end{equation}
where $\{t_1,...,t_k\}$ are the \textit{non-reuse} axes and the coalesced coordinates are given by the mixed-radix linearization
\begin{equation}
r = \sum_{j=1}^{q} i_j^{\text{out}} \prod_{j'=j+1}^{q} s_{j'}^{\text{out}}, \qquad c = \sum_{j=1}^{p} i_j^{\text{in}} \prod_{j'=j+1}^{p} s_{j'}^{\text{in}}.
\end{equation}
For the example in Fig.~\ref{fig:computation-mapping}(b), axes $i$ and $j_i$ are coalesced into $q$ for input reuse, while $p$ requires no coalescing as it already forms a single \textit{output-reuse} axis.

\textbf{Post-Tiling Schedule}.
Since physical macros have finite dimensions $R \times C$, iterations exceeding these bounds must be tiled by the post-tiling schedule:
\begin{equation}
\mathcal{S}_{\text{P\_tile}} = \{ (t_1,...,t_k, r, c) \rightarrow (t_1,...,t_k, r_o, c_o, r_i, c_i) \},
\end{equation}
where $(r_i, c_i) = (r \bmod R, c \bmod C)$ map directly to macro rows and columns, while $(r_o, c_o) = (\lfloor r/R \rfloor, \lfloor c/C \rfloor)$ form outer loops over macro-sized computation blocks.
Fig.~\ref{fig:computation-mapping}(c)-(d) shows the final mapping for our Conv1D example, with $p_i$ and $q_i$ bound to $r_i$ and $c_i$, respectively.
With data reuse now aligned to macro structure, remaining iterations can be distributed temporally or across multiple macros.

\subsubsection{Multi-macro Mapping}

While single macro mapping maximizes local utilization, modern CIM architectures often contain multiple macros to accommodate more complex workloads.
However, the selection of parallelization strategies is not always straightforward, as different choices may lead to vastly different data movement costs.
PolyCIM employs a priority-based distribution strategy that maps outer loops across macros based on their reuse patterns.
It first maps \textit{output-reuse} axes across macros, since partial sums typically have larger bit-widths than inputs, making their movement and storage particularly costly.
The strategy then maps the \textit{input-reuse} axes, allowing multiple macros to share the same input broadcasts to reduce data movement across the memory hierarchy.
\textit{Non-reuse} axes are mapped subsequently to provide additional parallelism without data dependencies.
Finally, \textit{weight-reuse} axes enable weight replication and are considered only when sufficient storage capacity is available across macros.
The mapping iteratively assigns axes to the macro dimension in the above order, splitting axes when they exceed the available macro count.

\subsection{Data Movement Optimization}

The polyhedral transformations that expose reuse opportunities fundamentally alter memory access patterns, creating a mismatch between original tensor layouts and the transformed iteration space. 
PolyCIM addresses this through two complementary optimizations: layout transformation that ensures contiguous memory access, and data movement scheduling that minimizes transfers across the memory hierarchy.

\subsubsection{Iteration-Space-Aligned Layout}
After transformation, the original data layout (e.g., channel-first or channel-last ordering for convolutions) no longer provides contiguous access, resulting in scattered memory requests that underutilize bandwidth. 
We introduce an \textit{iteration-space-aligned} layout that reorganizes data to match the transformed access order. 
For an array $T$ with a transformed access relation $\mathcal{A}_T$, we construct $T_{\text{align}}$ such that consecutive iterations access consecutive memory locations:
\begin{equation}
T_{\text{align}}[i_{p_1}, \ldots, i_{p_k}] = T[\mathcal{A}_T(i_1, \ldots, i_n)],
\end{equation}
where $\{i_{p_1}, \ldots, i_{p_k}\}$ form a basis for the accessed subspace of $T$.
This layout transformation is performed offline for all operands before computation begins. 
PolyCIM automatically generates the transformation code based on the specific polyhedral schedules applied, ensuring data is properly reorganized without manual intervention.

\begin{figure*}[t]
    \centering
    \includegraphics[width=\textwidth]{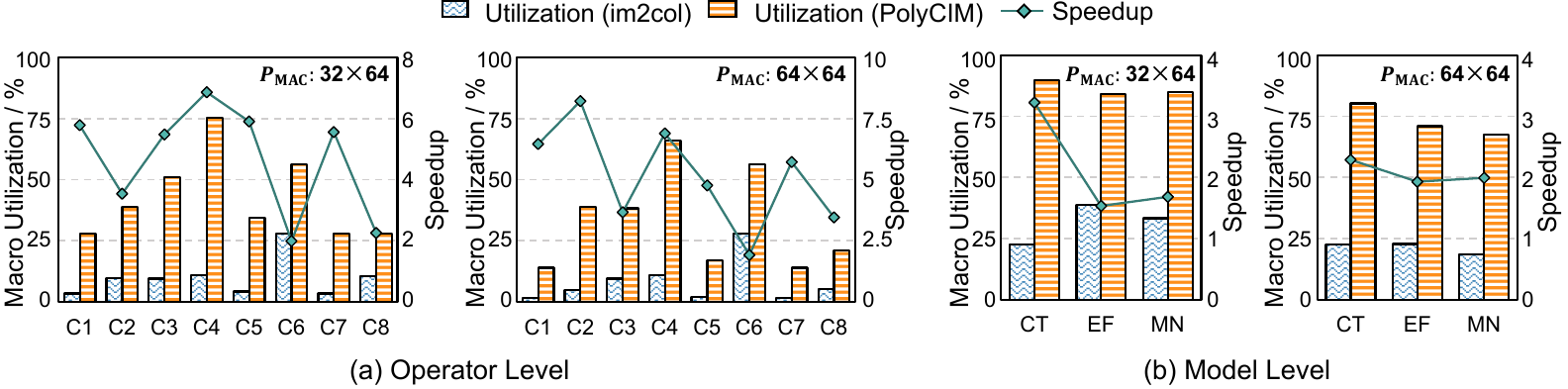}
    \caption{Performance comparison across different operators and models.}
    \Description{Four charts compare im2col and PolyCIM macro utilization with bars and show PolyCIM speedup with a line. The operator-level charts cover C1 through C8 on 32-by-64 and 64-by-64 macros, and the model-level charts cover ConvNeXt-Tiny, EfficientNet-B0, and MobileNetV2 on the same macro shapes.}
    \label{fig:performance-operator-network}
\end{figure*}

\subsubsection{Data Movement Scheduling}
Beyond layout transformation, PolyCIM optimizes when and where data moves through the memory hierarchy by jointly determining loop ordering and buffering points. 
Specifically, the optimization decides which permutation of loops minimizes transfers and at which loop level each operand should be fetched into buffers.

We formulate this as an integer optimization problem with two sets of decision variables.
A permutation matrix $\mathbf{X} \in \{0,1\}^{m \times m}$ represents loop reordering, where $X_{i,j} = 1$ means that original axis $j$ moves to new position $i$.
For each operand, a vector $\mathbf{L} \in \{0,1\}^m$ marks the buffering point, with $L_i = 1$ indicating loop levels that iterate against cached data and $L_i = 0$ indicating outer levels traversed before the data is cached.
The formulation leverages the insight that both buffer usage and transfer counts factor as products over the loop nest: the buffer size is the product of operand sizes along axes inside the buffering boundary, while the transfer count is the product of loop bounds along axes outside it.
We apply log-linearization to both quantities, which turns the multiplicative buffer-size and transfer-count expressions into sums.
For instance, data movement volume for a single operand, which equals buffer size times transfer count, can be formulated as
\begin{equation}
\begin{aligned}
    \log V &= \log(B \cdot T) = \log B + \log T \\
    &= \sum_{i,j} X_{i,j} \cdot (L_i \log A_j + (1-L_i) \log S_j),
\end{aligned}
\end{equation}
where $A_j$ represents the memory footprint and $S_j$ the loop bound for axis $j$.
The solver minimizes the sum of log-transformed data-movement volumes across all operands and memory levels while respecting capacity constraints. 
Despite its non-linear appearance, the formulation has only binary variables and a typically small problem size, allowing it to be solved efficiently using modern mixed-integer optimization solvers.

\section{Evaluation}\label{sec:evaluation}

\subsection{Experimental Setup}

We implement PolyCIM in Python using ISL for polyhedral optimization and the Gurobi solver for mixed-integer optimization.
While PolyCIM is architecture-agnostic and can target various CIM backends, we validate our approach by generating domain-specific language IRs for the CIMFlow~\cite{qi-2025-cimflow} framework, a CIM compilation and simulation infrastructure with complete instruction set architecture support.
We conduct experiments on an 8-macro CIM architecture with two representative SRAM-based digital CIM macro designs, with the concurrently activated rows and columns ($P_\text{MAC}$) being $32 \times 64$~\cite{yan-2022-sram} and $64 \times 64$~\cite{chih-2021-sram}.
We evaluate PolyCIM against the im2col-based mapping in CIMFlow, as specialized methods like SDK do not support our benchmark operators.
As detailed in Tab.~\ref{tab:benchmark_operators}, our experiments are conducted with operators from various DNNs, which cover different types and sizes to ensure comprehensive evaluation. 
To evaluate the end-to-end performance benefits, we also perform experiments on three complete DNN models with modern operators, including ConvNeXt-Tiny (CT), EfficientNet-B0 (EF), and MobileNetV2 (MN).
We measure both macro utilization and execution speedup through cycle-accurate simulation.

\begin{table}[t]
    \centering
    \caption{Operators used in performance evaluation.}
    \label{tab:benchmark_operators}
    \resizebox{\linewidth}{!}{%
        \begin{tabular}{llll}
            \toprule
            \textbf{Name} & \textbf{Type} & \textbf{Size} & \textbf{From} \\
            \midrule
            
            C1 & \multirow{5}{*}{DWConv} & $c=32,k_h=k_w=3$ & EfficientNet-B0 \cite{tan-2019-efficientnet} \\
            C2 &  & $c=240,k_h=k_w=5$ & EfficientNet-B0 \\
            C3 &  & $c=192,k_h=k_w=7$ & ConvNeXt-Tiny \cite{liu-2022-convnext} \\
            C4 &  & $c=1024,k_h=k_w=13$ & RepLKNet-31B \cite{ding-2022-replknet} \\
            C5 &  & $c=32,k_h=1,k_w=11$ & InceptionNeXt \cite{yu-2024-inceptionnext} \\
            \midrule
            C6 & GConv & $c=128,k=3,g=32$ & ResNeXt-50 \cite{xie-2017-resnext} \\
            \midrule
            C7 & DILConv & $c=128,k=3,d=2$ & ESPNetv2 \cite{mehta-2019-espnetv2} \\
            \midrule
            C8 & 3DConv & $k_h=k_w=k_z=3$ & MobileStereoNet \cite{shamsafar-2022-mobilestereonet} \\
            
            \bottomrule
        \end{tabular}%
    }
\end{table}

\subsection{Performance Results}
\subsubsection{Operator-Level Evaluation}

Fig.~\ref{fig:performance-operator-network}(a) shows the speedup and macro utilization achieved by PolyCIM compared to the im2col-based mapping approach, where PolyCIM consistently outperforms the baseline across all the evaluated convolution operators.
PolyCIM achieves speedups ranging from $1.9\times$ to $8.2\times$, with an average speedup of $4.9\times$.
The results also show that the rigid im2col mapping achieves particularly poor utilization (often below 10\%) for operators with non-standard reuse patterns, while PolyCIM obtains up to 75\% utilization through affine transformations to improve data reuse.
On the $32\times 64$ configuration, PolyCIM utilization scales cleanly with kernel size across the depthwise operators, rising from 28\% at $k{=}3$ (C1) to 75\% at $k{=}13$ (C4), as larger kernels offer more non-axial reuse for the affine transformation to expose.
The same consistency carries over to the $64\times 64$ configuration, where PolyCIM still improves on the im2col baseline for every operator and delivers the single highest speedup of $8.2\times$ on C2.
The absolute utilization ceilings fall slightly on this larger macro, most visibly for C4 which drops from 75\% to 66\%, likely because the additional rows are harder to fill productively once the operator dimensions no longer evenly divide them.
Grouped convolution (C6) and 3D convolution (C8) see smaller relative gains of roughly $2\times$.
For these operators, the broadcast alignment used by im2col already fits the reuse pattern reasonably well, so there is less idle capacity for PolyCIM to reclaim.
This demonstrates that exposing non-axial reuse opportunities through affine scheduling can effectively address the CIM utilization bottleneck.

\begin{figure}[t]
    \centering
    \includegraphics[width=\linewidth]{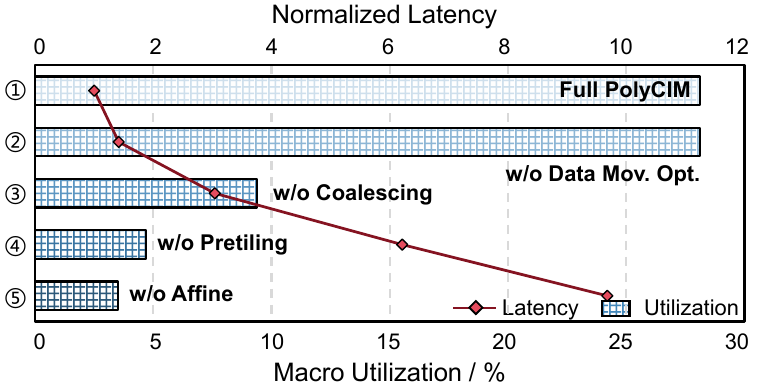}
    \caption{Performance impact of each optimization technique on the C1 operator.}
    \Description{An ablation chart compares full PolyCIM with variants that remove data-movement optimization, coalescing, pre-tiling, or affine scheduling. Horizontal bars show macro utilization, and a connected line shows normalized latency.}
    \label{fig:ablation-study}
\end{figure}

\subsubsection{Model-Level Evaluation}

Fig.~\ref{fig:performance-operator-network}(b) illustrates the end-to-end speedup and average macro utilization improvement across three modern DNN models.
PolyCIM achieves an average speedup of $2.1\times$ across all three models, with the primary acceleration stemming from optimized DWConv operations.
ConvNeXt-Tiny exhibits the highest speedup at $3.2\times$, due to its larger proportion of DWConv computations.
The utilization improvements further highlight the effectiveness of our approach.
On the $32\times 64$ configuration, compared to the $20\% \sim 40\%$ macro utilization of im2col mapping, PolyCIM reaches $84\% \sim 89\%$ utilization across all DNN models.
These model-level utilization figures exceed the per-operator numbers above because full networks still contain many standard convolution layers that im2col maps efficiently, and averaging those well-supported layers in with the emerging operators raises the overall utilization.
These results demonstrate that the evolution of DNN architectures toward efficient operators like DWConv has inadvertently created a fundamental mismatch with the structural constraints of CIM accelerators, which PolyCIM resolves through systematic polyhedral optimization.

\subsection{Ablation Study}

To quantify the contribution of each optimization component, we conduct an ablation study with the C1 operator by selectively disabling individual techniques.
We evaluate the following four key components: pre-tiling, affine scheduling, coalescing, and data movement optimization.
Fig.~\ref{fig:ablation-study} reveals the performance impact of each technique.
Without affine scheduling, latency degrades by $9.7\times$ as non-axial reuse patterns remain hidden, validating our core insight that these patterns are fundamental to modern operator efficiency.
Affine scheduling is the central step of the pipeline, as it is the only component that actually exposes new reuse, while the remaining optimizations refine the already-transformed iteration space.
Disabling pre-tiling causes a $6.2\times$ slowdown due to uncontrolled iteration space expansion after affine transformation.
The coalescing schedule proves essential for utilization, as its removal drops macro utilization from 28\% to below 10\% since scattered reuse dimensions often lead to underutilization.
Data movement optimization contributes a $1.4\times$ speedup by ensuring contiguous memory access after transformation.
These results demonstrate that while polyhedral transformation exposes reuse opportunities, realizing performance gains requires the complete optimization pipeline to handle the cascading effects of iteration space transformation.

\section{Conclusion}\label{sec:conclusion}

In this paper, we propose PolyCIM, a polyhedral-based compilation framework that addresses the fundamental mismatch between modern DNN operators and digital CIM architectures.
We identify that data reuse in DNNs forms hyperplane structures along non-axial directions, and demonstrate how polyhedral transformations can systematically expose these hidden reuse opportunities.
Through a unified polyhedral abstraction and a targeted optimization pipeline, PolyCIM can represent diverse DNN workloads and align data reuse opportunities with the rigid constraints of CIM array structures. 
Our extensive evaluation demonstrates the effectiveness of PolyCIM across diverse DNN workloads, achieving up to $4\times$ macro utilization improvement and $3.2\times$ speedup.

\bibliographystyle{ACM-Reference-Format}
\bibliography{ref}

\end{document}